\documentclass[12pt,letterpaper]{article}
\usepackage{amsmath,amssymb,pgf,pgfarrows,pgfnodes,float,appendix, hyperref}
\usepackage{graphicx}
\usepackage{subfigure}
\usepackage[margin=0.9in]{geometry}

\newcommand{\be}{\begin{equation}}
\newcommand{\ee}{\end{equation}}
\newcommand{\bea}{\begin{eqnarray}}
\newcommand{\eea}{\end{eqnarray}}

\title{{\rm\footnotesize \qquad \qquad \qquad \qquad \qquad \ \qquad \qquad \qquad \ \ \ \ \ \                 RUNHETC-2023-39}\vskip.5in   On the Impossibility of Precise Verification of Models of Quantum Gravity}
\author{Tom Banks\\
Department of Physics and NHETC\\
Rutgers University, Piscataway, NJ 08854\\
E-mail: \href{mailto:tibanks@ucsc.edu}{tibanks@ucsc.edu}
\\
\\
\\
\\
}
\date{}
\begin{document}
\maketitle

\begin{abstract} We argue that no theoretical model of quantum gravity in a causal diamond whose boundary has finite maximal area, can be verified with arbitrary precision by experiments done in that diamond.  This shows in particular that if our own universe remains in an asymptotically future de Sitter state for a time long enough for our local group of galaxies to collapse into a black hole, then no information processing system with which we can communicate could ever distinguish between many competing models of the AsdS universe.  This article is written in an attempt to be accessible to a wide audience, so certain elementary facts about quantum mechanics are reviewed, briefly.
\end{abstract}

\section{Classical Dreams and Quantum Measurements}

From the time that Newton and Leibniz invented calculus, the implicit goal of theoretical physics has been to construct a model that could, in principle, make infinitely precise predictions about the future state of the universe, given infinitely precise knowledge of its present state.  This was stated most succinctly in a famous sentence of Laplace
\begin{center}
{\it Given for one instant an intelligence which could comprehend all the forces by which nature is animated and the respective positions of the beings which compose it, if moreover this intelligence were vast enough to submit these data to analysis, it would embrace in the same formula both the movements of the largest bodies in the universe and those of the lightest atom; to it nothing would be uncertain, and the future as the past would be present to its eye.}
\end{center}
Laplace was of course also one of the creators of the theory of probability, for he recognized the impossibility of actually knowing everything about the initial state of the universe with sufficient precision to make accurate prediction possible.   The utility of the theory of probability rests on an assumption, whose mathematical statement is that the probability of a system going from state A to state B in time $t$ is the sum of the probabilities of all possible histories by which the system could have gotten between A and B in time $t$.  It's this rule that allows the weatherperson to make more accurate predictions about the future track of a hurricane, after they know whether it has hit New Orleans or Galveston on a particular day.  Their equations predicted similar probabilities for both events.  

Quantum mechanics throws a wrench into this scheme for precision prediction limited only by the precision of one's knowledge of the present.  QM does not obey the sum over histories rule for probabilities.  This rule is so embedded in our ordinary experience that we consider it part of "logic" and all of the confusion about the foundations of quantum mechanics has to do with the fact that it violates the sum over histories rule.   

It was understood intuitively by Bohr and Heisenberg, and on a much more technical level by at least {\it some} quantum physicists since the 1970s, that the essence of "quantum measurement theory" was the fact that certain quantum systems have a large variety of {\it collective variables} $C_i$.  These variables have two interesting properties.
They're defined as averages of local variables over volumes that are "large in microscopic units" .   The quantum statistical uncertainties in these variables are of order the inverse square root of the large volume.  Even more important, the violation of the sum over histories rule for the probabilities of these variables is {\it exponentially small as a function of the large volume}.   To get a feeling for what we mean by large volume, if we talk about a cube that's one tenth of a centimeter on each side, then the quantum uncertainties are of order $.0000000001$ and the violations of the sum over histories rule are of order $10^{-100000000000000000000}$.   We've had to use exponential notation for the last number because if we wrote it out in decimal form on $8\times 10$ sheets of paper in a normal font it would take a stack of pages from here to the planet Saturn to fit it in. 

Such collective variables appear very naturally in quantum systems that are composed of lots of individual variables at independent points of space.  It's convenient to think of space as a very fine grid of points with independent variables at each one.  We'll return to the question of whether this is a good model of what space is really like, but since the middle of the 19th century, physics has been based on models like this, which are called {\it field theories}.   Field theories naturally have lots of collective variables defined as averages of fields over many points.  Quantum mechanical field theories are the basis of the standard model of particle physics, which accounts for all known experimental data within the accuracy of theoretical computation and experimental precision\footnote{We're assuming that we've added terms to the standard model to account for neutrino masses and possibly other terms to account for recent discrepancies between theoretical and experimental values of the magnetic moment of the muon. These terms fall within the well understood formalism of quantum field theory.}

In the 19th century, there were three field theory models know to theoretical physicists: Maxwell's theory of electromagnetism, Newton's theory of gravitation, and the theory of hydrodynamics.  Hydrodynamics had many fathers and should really be seen as encompassing the motion not only of liquids, but the theory of elasticity in solids and the propagation of sound.  It turned out that Maxwell's theory was a fundamental quantum theory, while hydrodynamics was a very universal phenomenological theory describing the propagation of long wavelength disturbances in any kind of matter.  In most circumstances, the quantum behavior of the matter was not properly described by applying the rules of quantum mechanics to the equations of hydrodynamics\footnote{This {\it is} a proper way to treat the very low energy excitations of the ground states of many quantum systems.}.   

The complete field theory model of gravitation, General Relativity (GR), was discovered in 1916 by Albert Einstein and it introduced an entirely new feature into the story.  In all previous theories, the geometry of space was that of Euclid.  Even Einstein's revolutionary Special Theory of Relativity (1905) did not change that.  It only proposed that the description of spatial geometry used by systems in relative motion differed by a scale factor.    GR says that the spatial geometry is generally non-Euclidean ({\it i.e.} curved), responds to the matter embedded in it, and changes with time!   

One way of describing the geometry of space in GR, which uses Einstein's principle that nothing can travel faster than the speed of light, $c$, is to imagine some sort of information gathering system traveling through it, in such a way that at each time its velocity is less than that of light.  The system has a clock on it, which measures what we call its {\it proper time}.  In any given interval of proper time light can only have traveled out as far as some maximally distant surface, and we record the area of that surface for each interval of time.   Do this for all possible intervals of time and all possible information gathering systems, and you've completely determined the geometry of space for all time.  

The strange dynamical geometry of space shows up in the simplest possible non-trivial solution of Einstein's gravitational field equations: the analog of the Newtonian gravitational field of a point mass.  This solution was first found by Schwarzschild a few years after Einstein published his field equations, but was not properly understood until the 1960s.   Newton's gravitational constant $G_N$ and the mass $M$ of an object define a length scale, $G_N M$.   $2G_N M = R_S $ is now called the Schwarzschild radius of an object of mass $M$.  Schwarzschild found that outside the Schwarzschild radius one could choose coordinates for space and time such that the spatial geometry is static and gives rise to the Newtonian potential at large distances from the center.  Inside the Schwarzschild radius the spatial geometry is rapidly time dependent.  An invariant way to characterize what is going on is again to look at an arbitrary information gathering system that falls through that radius.   It cannot send a light signal out to a different system that remains outside the Schwarzschild radius.  However one defines the "space inside" it is expanding away from the Schwarzschild radius "faster than the speed of light".  Secondly, two different information gathering devices thrown in to the Schwarzschild radius at the same time but at different angles, can meet only if they do so in a time less than $R_S/c$.  Another way to say this is that as the clock on any of those interior systems ticks away, the area of the surface that it can explore by sending out light rays and getting back their reflection, shrinks to zero in a time about $R_S/c$.   

If we have a star of mass $M$ and radius $R \gg R_S$ then the interior Schwarzschild region is buried inside the matter of the star and the simple Schwarzschild solution does not apply.  However, work beginning with that of Tolman and Oppenheimer and Volkoff and culminating in a tour de force paper by Chandrasekhar, showed that any sufficiently massive star would have a similar "black hole region", which would swallow up the whole star.  In general the star has angular momentum and it could have non-zero charge, so one needs a more general solution of Einstein's equations than Schwarzschild's, but those solutions have similar properties.

\section{Quantum Theories of Gravitation}

Einstein's GR taught us that, in contrast to Maxwell's theory of electromagnetism, which was viewed as a model of waves moving a fixed space-time, the theory of gravitation made space-time dynamical.  This disparity was removed by Kaluza and Klein, who showed that electromagnetism could be the consequence of dynamical geometry in $4 +1$ space-time dimensions, if the fourth spatial dimension was a circle whose radius always remained very small in normal units.  Modern string theory models have shown that in principle the entire structure of the standard model of particle physics plus Einstein's GR, could be a consequence of dynamical geometry in $10 + 1$ dimensions. No model that precisely fits the standard model has yet been found in string theory, but the list of possible models is far from complete and the existing list is vast and contains many examples that come very close to reality.  This makes the question of how to "quantize" GR the central question of high energy theoretical physics.

As a first step in thinking about this question we should talk about units in physics.  If you've ever taken an elementary physics class you've been bewildered by all the names of units for different physical quantities, energy,mass,temperature,electric field, magnetic field, length, time, and so on.  This confusion is historical and reflects our initial ignorance about how different things were connected together.   With the advent of Einstein's theories of relativity, everything was reduced to a single unknown unit, a unit of length.  When Max Planck introduced his famous formula for the spectrum of radiation from a hot oven, which signalled the discovery of quantum mechanics, he considered that one of the most important aspects of it was the introduction of his fundamental constant $\hbar$, which gives a minimal energy for a given frequency of light, because together with Newton's gravitational constant and the speed of light it defines a fundamental unit of length.  Newton's constant, in "natural units" ,where $\hbar = c = 1$, has the dimensions of an area $G_N = L_P^2  = 10^{-66}$ cm$^2$.  


Now we can ask the fundamental questions: {\it What are space and time?}.  As emphasized by J.L. Borges in a paradoxically entitled essay\cite{borges} one cannot say a sentence in any human (or computer) language without implicitly referring to the passage of (proper) time.  Einstein taught us that time is relative.  Different information gathering systems may have different measurements of how time passes when viewing the same set of information.  But for any given system, proper time is a primitive concept without which we (or the system) can't express a thought.  

Space, on the other hand, might be replaced by a more primitive concept, namely information.  J.A. Wheeler invented a clever motto for this idea "It from Bit", which has been updated to "It from q-bit".   A {\it bit} is the smallest amount of information one can think about, the answer to a Yes/No question, and a {\it q-bit} is the quantum version of a bit.  Alan Turing, the genius who broke the German Enigma code, realized that any finite set of information could be encoded in some number of bits.  For example, $2$ bits have $4$ possible states, but if we make the rule that we don't allow the state where both answers are Yes, then there are only $3$.   In a similar way, any finite set of possible answers can be thought of as answers to a bunch of independent Yes/No questions, with {\it a priori} constraints that certain combinations of Yes answers are not allowed. 

To get an idea of what a q-bit is draw a picture where an arrow of length $1$ pointing up represents Yes, and an arrow pointing to the right represents No.  
\begin{figure}[h]
\begin{center}
\includegraphics[width=01\linewidth]{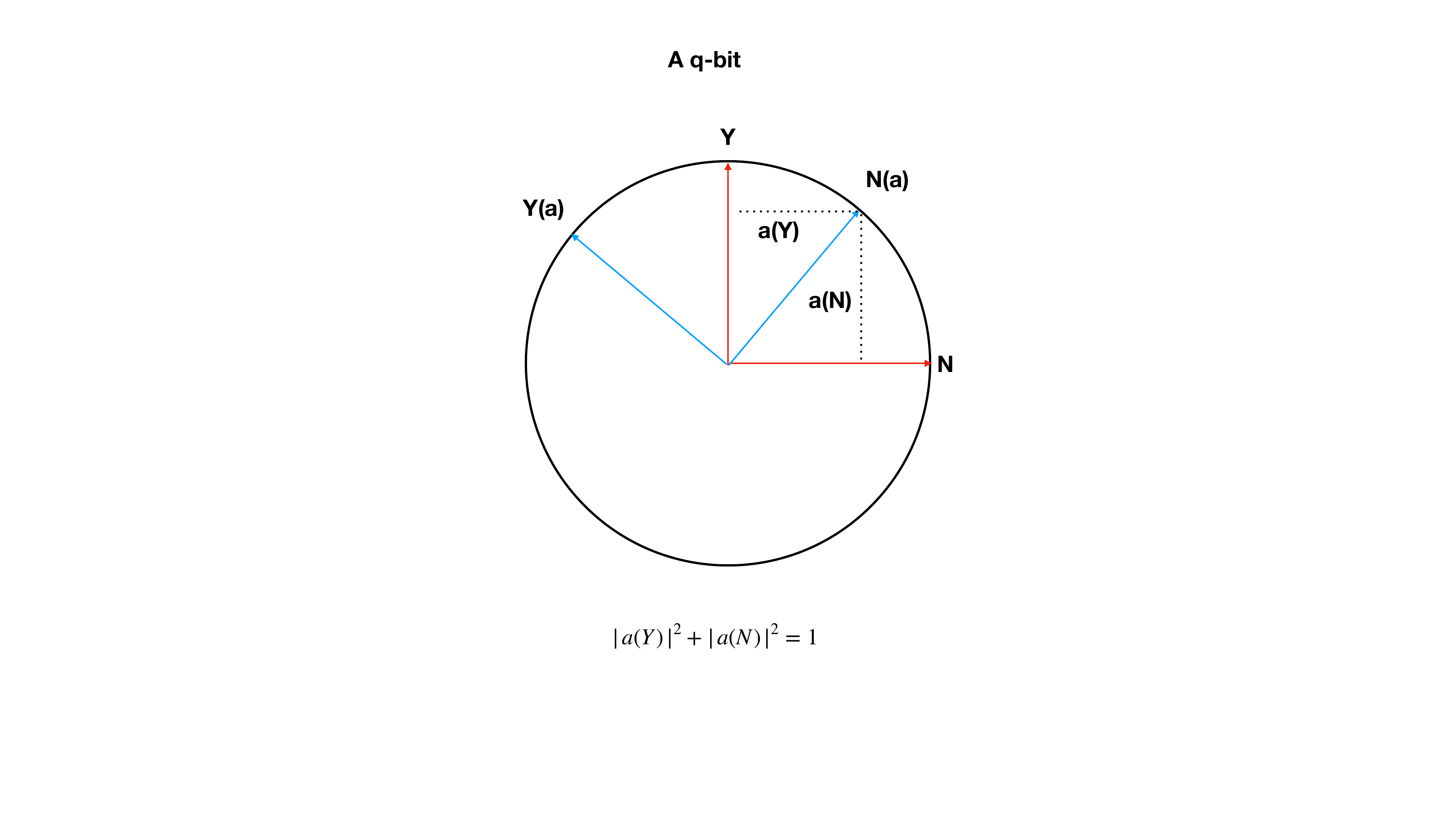}

\caption{Mnemonic for a q-bit. Different orientations of the axes represent different measurements that can be done on the q-bit, with the absolute squares of the projections of one set of axes on the others giving the probabilities that one set of measurements will have particular answers when the answers to the other set are definite.} 
\label{aq-bit}
\end{center}
\end{figure}

You can think of the arrow as being the lever on a valve in a water pipe, with the up direction being the direction that lets the water flow and the sideways direction the one that blocks it.  Now draw a picture with the two arrows rotated by an angle $a$.  If you remember your high school trigonometry, the projection of the new Yes direction on the old one is $\cos a$ and on the old No direction is $\sin a$ and
\begin{equation} \sin^2 a + \cos^2 a = 1 . \end{equation}
In QM this is interpreted as the existence of a new state of the system, in which the probability of the answer to the original question being Yes is $\cos^2 a$.   A q-bit is just the statement that we allow all of these new states with arbitrary angles, so that not every question has a definite answer.   It's actually a little more complicated than that: we have to introduce complex numbers, but it would take us too far afield to explain that.   Since you're reading this archive, I'm going to assume that you know enough about QM to go on and that further explanation would just bore you.  Basically a q-bit is just a bit that can be looked at in many ways that are mutually incompatible.  When one version of the q-bit's question is answered with absolute certainty, then the system is in a state where one can only make probabilistic predictions about what the result of a "measurement" of any of the different versions of the question will be.  In order to make those measurements and verify the probabilistic predictions we have to make repeated correlations of the q-bit system with a collective variable of some much larger quantum system and record the frequency with which we get Yes and No answers.

In the real world, q-bits or quantum information are carried by physical systems.  For example, the spin of an electron can be a q-bit.  So it makes sense to ask whether there's a maximum number of q-bits that fits into a certain region of space.  This is a modern version of the medieval question of how many angels can dance on the head of a pin.  We can actually turn the question around though and {\it define space} by the number of q-bits that fit into it.  We do this using the description of GR that we mentioned at the end of the previous section.  For every information gathering system in a space-time and every proper time interval $T$ on its clock,  we have a pure number $A(T)/4G_N$ where $A(T)$ is the area of the maximally distant surface that the system could have explored by bouncing light beams off of it.  We call the region of space-time explorable by the system the {\it causal diamond of the experiment}. For more or less flat space like that near us, the area of the causal diamond is about $4\pi T^2$.   This pure number should be related to the number of q-bits accessible to the system.    More q-bits, more area.  Jacob Bekenstein first conjectured a formula like this based on the properties of black holes and the laws of thermodynamics.

Stephen Hawking had shown that the total area of black hole horizons in the universe always grew with time, just like the quantity called "entropy" in thermodynamics.  Entropy had been shown by Boltzmann in the 19th century to be the amount of information hidden in complex systems, which led to apparent violations of the law of conservation of energy by processes like friction.  The energy doesn't disappear, but goes into complex motions of the microscopic constituents of the systems, which we perceive only as "heat".   More modern investigations have revealed that entropy counts the logarithm of number of accessible quantum states of the microscopic constituents in a given set of macroscopic conditions.  Hawking, who initially dismissed Bekenstein's conjecture, showed that black holes had a temperature, and computed the coefficient in the entropy formula.  He found that the entropy was exactly $A/4G_N$. In 1995, Jacobson showed\footnote{Because of certain mis-statements in current literature, I feel compelled to insert a fairly lengthy footnote here. The area law for entanglement entropy of a causal diamond was first written down by Sorkin in 1983\cite{sorkin} and rediscovered by Srednicki\cite{sred} and Callan and Wilczek\cite{cw}.  This led Susskind and Uglum\cite{sussug} and Jacobson\cite{jacob} to make the conjecture that this was somehow related to the renormalization of Newton's constant in the Bekenstein-Hawking area law for black holes. No one commented on the revolutionary leap being made, since there were no black holes in sight, but Jacobson surely understood because he soon showed that the hydrodynamics of this law was equivalent to Einstein's equations doubly projected on arbitrary null vectors\cite{ted95}. Jacobson's paper was written in terms of small changes in the size of a causal diamond, so he never made the explicit Covariant Entropy Conjecture.  That was first made for cosmological space-times by Fischler and Susskind in 1998 and for general space-times by Bousso in 1999.  This led Fischler and myself\cite{tbwf}, independently, to postulate that the density matrix of empty dS space was the unit matrix on a finite dimensional Hilbert space, with dimension determined by the Gibbons Hawking entropy formula.  When we later extended this hypothesis to the general CEP, Bousso pointed out that he had made the same conjecture in one of his big reviews on the Holographic Principle in 1999.  The most important consequence of this observation, that localized states in dS reduce the entropy, giving an explanation of the dS temperature, was something Fischler and I recognized immediately, but which did not get put into print until B. Fiol showed me how to make an explicit quantum mechanical model of the effect in 2006\cite{bfm}.} that the {\it Covariant Entropy Principle} (CEP)implied all of Einstein's equations except the so-called cosmological constant term.  The CEP states that the Bekenstein-Hawking relation between area and entropy holds for {\it every} causal diamond in every space-time, not just the causal diamonds of systems that have fallen inside a black hole horizon.

The CEP is in tension with the mathematics of quantum field theory.  The standard model of particle physics, or any other quantum field theory, would tell us that the logarithm of the number of quantum states that could fit into the causal diamond of an experiment done over proper time $T$, scales like $(T/L)^3$, where $L$ is the shortest wavelength we allow in the fields.  It's plausible that $L$ is about $L_P$. On the other hand, most of those states have very high energy and high energy creates strong gravitational fields, which means black holes.   If we throw away states that would have created black holes with area larger than about $(T/L_P)^2$, then the log of the number of states is cut down to $(T/L_P)^{3/2}$, {\it which is much less than the entropy implied by the CEP}.   

In 1998, Cohen, Kaplan and Nelson\cite{CKN}, showed that one can omit all of the states that would have created large black holes from quantum field theory, without having any detectible effect on the most precisely known agreement between quantum field theory and experiment.   So it's extremely likely that, whatever the theory of quantum gravity in the region accessible to the information gathering device is, only a tiny fraction of its quantum states are described by quantum field theory.  The rest are black holes.

A fundamental insight into the nature of black hole quantum mechanics appeared in several publications by Lindesay, Susskind, Hayden, Preskill and Sekino\cite{lshpss}.  Susskind and Sekino gave the phenomenon the name of {\it fast scrambling of quantum information}.  It basically has to do with the fact that perturbations of a black hole disappear exponentially rapidly, leaving over only the macroscopic information about the hole's charge, mass and angular momentum.  Hydrodynamic flows on the black hole horizon are incompressible, which means that there is no propagation of information.   This means that although black holes, like quantum field theories, have many quantum states, they are not good information processing or storage devices.   Averages of quantities over part of the black hole horizon will, almost all of the time, just be fractions of the charge, mass, and angular momentum of the black hole.  The system does not have a complex set of collective variables that can measure the properties of a microscopic quantum system.

The final piece of our story is the discovery of what is called {\it the accelerated expansion of the universe}.  The simplest way to explain this is to add a positive {\it cosmological constant} to Einstein's equations.  Recall that the value of the cosmological constant was the one term that couldn't be determined from Jacobson's demonstration that the equations followed from local variations of the BH area law for general causal diamonds.  This is because the cosmological constant controls the relation between the limits of large proper time and large area. It is not a local energy density. When it is positive, the area remains finite as proper time goes to infinity, while if it's negative the opposite is true.  If it's exactly zero then they go to infinity together, with $A \sim T^2$.  Our universe appears to be approaching a so called de Sitter universe with a maximal radius $R$ about $10^{61} L_P$.  

We've now come to the fundamental conundrum of a theory of quantum gravity in a finite de Sitter universe, with radius $R$.   No matter how long an information gathering system exists, the total amount of information accessible to it is finite, but the total number of {\it useful} q-bits in which it can store and process that information is smaller by a factor $(R/L_P)^{-1/2} \sim 10^{-30}$ than that accessible information.  The number of semi-classical collective variables which can actually make reliable records of that information is smaller still.  Thus, there is a limit, {\it in principle} to the accuracy with which the information processing system can check any particular mathematical model of the entire system.  {\it A fortiori} a model based on infinite dimensional algebras is uncheckable because this requires an infinite number of measurements.  

There have been many suggestions in the literature that de Sitter space is unstable and claims that stable de Sitter space poses paradoxes because of the recurrences that occur in finite systems.  As long as the instabilities take place on a long enough time scale (and changes in the system sufficient to avoid recurrence paradoxes certainly take place on a long enough time scale) they do not change the conclusions of this article\cite{soniaetal}.  The aim of theoretical physics is to make predictions about potential observations.  If the universe continues its present evolution for about $100$ times its current age, our local group of galaxies will become causally disconnected from the rest of the universe.  Some time after that, the local group will collapse into a black hole, and theoretically possible measuring devices in our causal diamond will have ceased to exist.   Unless there is a drastic change in the evolution of the universe before that time, quantum gravitational theorists will have to content themselves with imprecise, finite theories.  It would be a good idea to concentrate on things that can actually be compared to experiment/observation.   In\cite{Mcosmo}, Fischler and I suggested that finite time analogs of scattering amplitudes would be the correct observables for an asymptotically dS universe.  We did not appreciate at the time the extent to which these failed to exhaust the available quantum states in a dS universe. These observables will, according to the arguments presented here, be adequately explained by a quantum mechanical model with a finite number of q-bits, whose details can never be precisely verified.   Laplacian "dreams of a final theory" were always meant to be a goal that could only be reached asymptotically.  The dual constraints of quantum mechanics and black hole formation imply that even that asymptotic goal is out of reach.  An information gathering system that exists for a finite proper time cannot, {\it in principle} perform a precise experimental check of a quantum theory of all the quantum states with which it is in causal contact.  An IGS in a future asymptotically dS universe cannot perform such a check even if the IGS persists forever.

To conclude, for aficionadas of string theory, we should explain how what we have said is consistent with the existence of precise formulae for the quantum gravitational S matrix in perturbative string theory in asymptotically flat space, and non-perturbative formulae in AdS space.  If we think of asymptotically flat space as the limit of dS space, then it is clear that the horizon states have to be thought of as converging to states of arbitrarily soft massless particles.   This leads one to contemplate a formulation of scattering theory in which states with non-zero momentum are defined in terms of constraints that set zero momentum operators to zero in certain regions on the sphere at null infinity\cite{tbscatt}. 
The Hilbert space of the theory is infinitely larger than what is captured by the S matrix, but one hopes that the infinitely soft sector decouples, at least from inclusive cross sections with a total missing energy cutoff.  Above four dimensions this problem may not arise until one attempts to go beyond perturbation theory.  Models in AdS space which are derived as decoupling limits of brane configurations in asymptotically flat space can be explained in a similar fashion, though here the decoupling of soft physics is much more transparent.

\end{document}